\documentclass[reprint,aps,pra,twocolumn,superscriptaddress,showpacs,longbibliography]{revtex4-1}
\usepackage{CJK}
\usepackage{bm,amsmath,amssymb,amsfonts}
\usepackage{color}
\usepackage{dcolumn}
\usepackage{graphicx}
\usepackage{float}
\usepackage{hyperref}
\hypersetup{colorlinks=true,citecolor=blue,urlcolor=blue,linkcolor=blue}
\usepackage{ulem}
\usepackage{times}

\newcommand{\finished}{{\color{red}\checkmark}}
\newcommand{\ket}[1]{|#1\rangle}
\newcommand{\braket}[1]{\langle#1\rangle}

\begin{document}

\title{Density Shift of Optical Lattice Clock via Multi-Bands Sampling Exact Diagonalization Method}%

\author{Yan-Hua Zhou}
\affiliation{Department of Physics, and Chongqing Key Laboratory for Strongly Coupled Physics, Chongqing University, Chongqing, 401331, China}

\affiliation{Center of Modern Physics, Institute for Smart City of Chongqing University in Liyang, Liyang, 213300, China}

\author{Xue-Feng Zhang}
\affiliation{Department of Physics, and Chongqing Key Laboratory for Strongly Coupled Physics, Chongqing University, Chongqing, 401331, China}
\affiliation{Center of Quantum Materials and Devices, Chongqing University, Chongqing, 401331, China}
\author{Tao Wang}
\thanks{corresponding author: tauwaang@cqu.edu.cn}
\affiliation{Department of Physics, and Chongqing Key Laboratory for Strongly Coupled Physics, Chongqing University, Chongqing, 401331, China}
\affiliation{Center of Modern Physics, Institute for Smart City of Chongqing University in Liyang, Liyang, 213300, China}
\affiliation{State Key Laboratory of Quantum Optics and Quantum Optics Devices, Shanxi University, Taiyuan, 030006,China}
\begin{abstract}
Density shift plays one of the major roles in the uncertainty of optical lattice clock, thus has attracted lots of theoretical and experimental studies. However, most of the theoretical research considered the single-band and collective approximation, so the density shift of the system at higher temperatures can not be analyzed accurately. Here, we design a numerical algorithm that combines Monte Carlo sampling and exact diagonalization and name it as Multi-Band Sampling Exact Diagonalization (MBSED). The MBSED method considers the collision of atoms between multi-bands, so it can provide the density shift of an optical lattice clock with higher precision. Such an algorithm will benefit the numerical simulation of an optical lattice clock, and may also be used in other platforms of quantum metrology. In addition, through our numerical simulation, we also found that the density shift of Rabi spectrum is slightly non-linear with atom number.
\end{abstract}
\maketitle

\section{Introduction}
Time, as one of the fundamental quantities, plays an important role in physics. The precise measurement of time is key to the frontier researches including measurement of fundamental constants \cite{detectingPhysicsConstant}, detecting gravitational waves \cite{Gravitational_wave}, testing general relativity \cite{katori_np,redshift,redshift2} and searching for the dark matter \cite{dark_matter1,dark_matter2}. As one of the most accurate time-frequency measurement devices, the optical lattice clock (OLC) platform has made great progress due to the experimental efforts in manipulating alkaline-earth or alkaline-earth-like atoms. Recently, OLC of Ye's group in JILA has reached an ultra-stable level with a fractional frequency measurement uncertainty $7.6\times10^{-21}$, so that the gravitational redshift at millimeter-scale can be resolved \cite {redshift}.  

To suppress the quantum projection noise, a large number of atoms need to be prepared in the ground state at each optical lattice site \cite{DickeLimit}. However, the collisions between atoms can degrade the precision of clocks \cite{atom_c}, so the corresponding quantitative calculation becomes vital for quantum metrology. On the other hand, understanding the collisional behavior is also important for both quantum simulations \cite{twoorbital,kondo,heavy,SU6,SUN,CGandWCJ2021} and quantum information \cite{QubitRegister, QuantumComputing,Quantumlogic} based on OLC. In the early work \cite{dingbiao}, only the s-wave scattering was considered due to the inhomogeneous excitation. Then, people found that the p-wave scattering in fact dominates the density shift \cite{probing}, and proposed the inter-atomic collisional model which describes the deep OLC under low longitudinal (z) temperature (around $1\mu K$). In the calculation, only the lowest longitudinal band is taken into account because the longitudinal energy gap is much larger than the temperature effect. Meanwhile, under the collective approximation, the frequency shift is found to be linear with the atom density, and such prediction is checked by the experiments of Ye's group \cite{aquantum}. 

However, sufficient cooling can not always be guaranteed. Considering the rapid progress of the transportable\cite{transportable} and space OLC\cite{space_clocks} which might be difficult to reach the low temperature compared with the lab, it is important to extend the collisional model to multi-bands case, so that the density shift can be estimated with high precision. In this paper, we first derive the multi-bands collisional model of OLC. Then, we developed an advanced algorithm which combines Monte Carlo sampling and exact diagonalization together to calculate the density shift of OLC. Thus, we name it as Multi-Band Sampling Exact Diagonalization (MBSED). In comparison with the collective approximation, the MBSED method presents higher accuracy, especially at higher temperatures. Therefore, our research can provide guidance and a benchmark for OLC working at relatively higher temperatures.

The paper is organized as follows. In Sec. \ref{sec2}, we discuss the effective spin model which further takes into account the longitudinal multi-band effect of quantum many-body systems in OLCs. 
In Sec. \ref{sec3}, the algorithm of multi-band sampling exact diagonalization (MBSED) is discussed. In Sec. \ref{sec4}, we compare the MBSED with the collective approximation for both Ramsey and Rabi spectroscopy. At last, Sec. \ref{sec5} includes the conclusions and outlook. 

\section{effective spin model}\label{sec2}
We considered an OLC that uses the $^1$S$_0(\ket{g})\leftrightarrow ^3$P$_0(\ket{e})$ clock transition in nuclear spin-polarized $^{87}$Sr atoms. The atoms are loaded in a deep optical lattice constructed by laser with magic wavelength \cite{magicwavelength}, so that they feel the same lattice potential with inter-site tunneling strongly suppressed. Then, each site can be taken as a 3D slightly  anharmonic oscillator with $N$ atoms staying at the different motional bands. Considering the pairwise interaction between atoms through the s-wave and p-wave channels \cite{QubitRegister,twoorbital,manybodytreatment,Hami1,Hami3,Inealstic}, The Hamiltonian can be written as:    
\begin{align}
	\hat{H}_o=&\sum_a\int\hat{\psi}_a^\dagger(\bm{r})[-\frac{\hbar^2}{2m}\nabla^2+V_{ext}(\bm{r})]\hat{\psi}_a(\bm{r})d^3\bm{r}\nonumber 
	\\
	+&\frac{4\pi\hbar^2a_{eg}^-}{m}\int\hat{\psi}_e^\dagger(\bm{r})\hat{\psi}_e(\bm{r})\hat{\psi}_g^\dagger(\bm{r})\hat{\psi}_g(\bm{r})d^3\bm{r}\nonumber
	\\
	+&\frac{3\pi\hbar^2}{m}\sum_{\alpha,\beta}b_{\alpha\beta}^3\int[(\nabla\hat{\psi}_\alpha^\dagger(\bm{r}))\hat{\psi}_\beta^\dagger(\bm{r})-\hat{\psi}_\alpha^\dagger(\bm{r})(\nabla\hat{\psi}_\beta^\dagger(\bm{r}))]\nonumber
	\\ 
	&\cdot[\hat{\psi}_\beta(\bm{r})(\nabla\hat{\psi}_\alpha(\bm{r}))-(\nabla\hat{\psi}_\beta(\bm{r}))\hat{\psi}_\alpha(\bm{r})]d^3\bm{r}\nonumber
	\\
	+&\frac{1}{2}\hbar\omega_0\int[\hat\rho_e(\bm{r})-\hat\rho_g(\bm{r})]d^3\bm{r}\nonumber\\
	-&\frac{\hbar\Omega_0}{2}\int[\hat\psi_e^\dagger(\bm{r}) e^{-i(\omega_L t-\bm{k}\cdot\bm{r})}\hat\psi_g(\bm{r})+h.c.
	]d^3\bm{r}\label{OriginalHamiltonian},
\end{align}
where $\hat\psi_\alpha(\bm{r})$ is a fermionic field operator at position $\bm{r}$ for atoms with mass $m$ in electronic state $\alpha$=$g$ or $e$ while $\hat\rho_\alpha(\bm{r})=\hat\psi_\alpha^\dagger(\bm{r})\hat\psi_\alpha(\bm{r})$ is the corresponding density operator, and $a_{eg}^{-}$ is the scattering length describing collisions between two atoms in the anti-symmetric electronic state, $\frac{1}{\sqrt{2}}(\ket{ge}-\ket{eg})$. The p-wave scattering volumes $b_{gg}^3$, $b_{ee}^3$, $b_{eg}^3$ represent three possible electronic symmetric states ($\ket{gg}$, $\ket{ee}$, and $\frac{1}{\sqrt{2}}(\ket{eg}+\ket{ge})$) respectively. $\omega_L$ is the frequency of probing laser and $\bm{k}$ is its wave vector. $\omega_0$ is the atomic transition frequency and $\Omega_0$ is the bare Rabi frequency.

\begin{figure}[t]
	\includegraphics[width=0.45\textwidth]{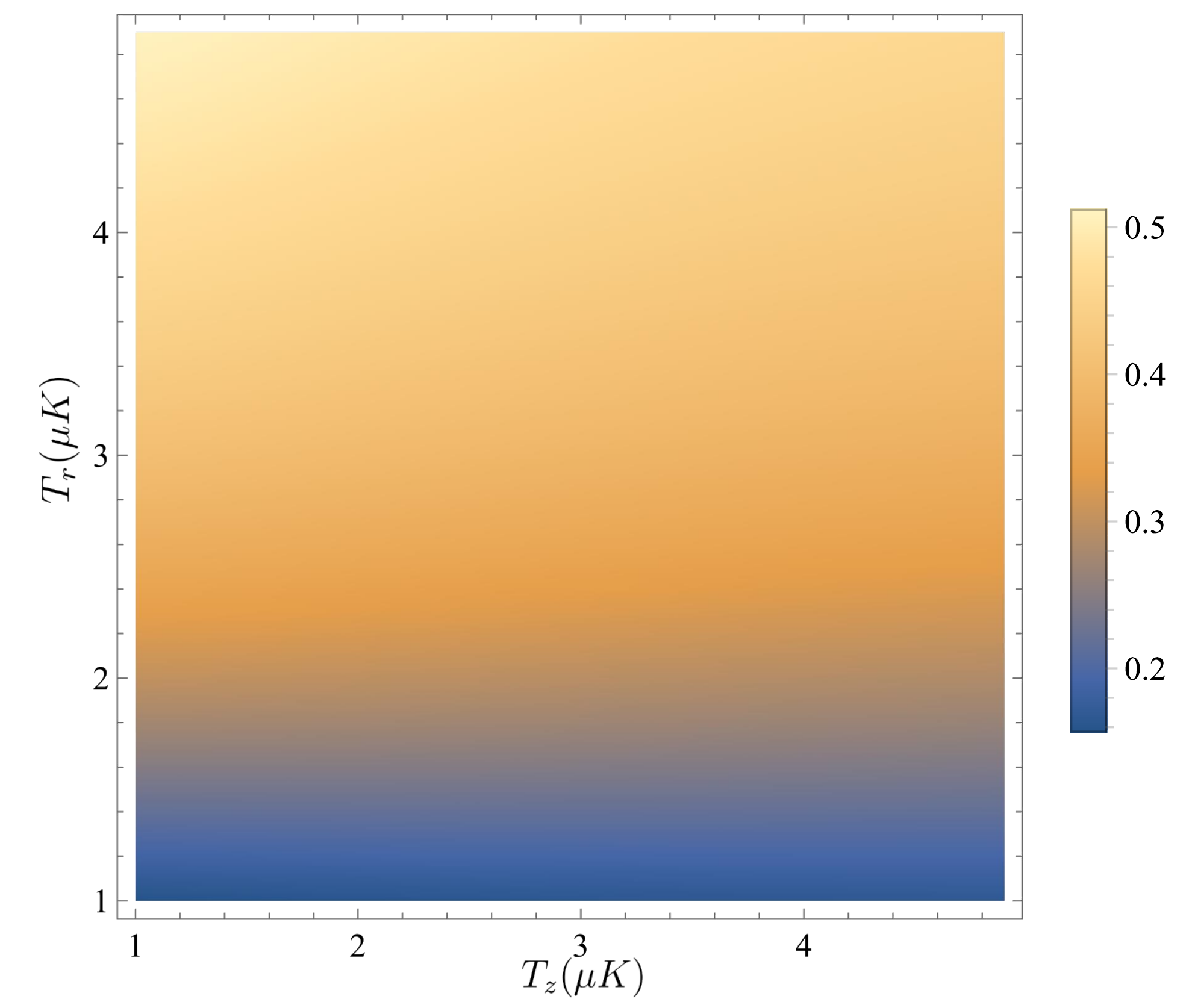}
	\caption{\label{fig1}The variation of Rabi frequency $\Delta\Omega/\overline{\Omega}$ as a function of $T_r$ and $T_z$, with the misalignment angle $\Delta \theta\approx10 $mrad.}
\end{figure}
The previous works \cite{aquantum,probing} focus on the system at the low temperature, so that only the lowest longitudinal band needs to be considered. In contrast, turning to the higher temperature,
the influence of higher longitudinal bands can not be ignored. Therefore, different from previous studies \cite{probing}, the expansion of the field operators in a non-interacting harmonic eigenbasis is not limited to the lowest longitudinal band :
\begin{equation}
	\hat{\psi}_\alpha(\bm{r})=\sum_{\vec{n}}\hat{c}_{\alpha\vec{n}}\phi_{n_x}(X)\phi_{n_y}(Y)\phi_{n_z}(Z), \label{basis}
\end{equation}
where $\hat{c}_{\alpha\vec{n}}$ annihilates a fermion in motional mode $\vec{n}=\{n_x,n_y,n_z\}$ and electronic state  $\alpha$.  Because all trap frequencies are much greater than the characteristic interaction energy, the motional degrees of freedom are effectively frozen \cite{probing}. Considering the energy conservation law and anharmonicity of the trap, the external states of atoms could only keep the same or exchange after the collision. Then the atom distribution in the external states is fixed and obtained by sampling as the Boltzmann distribution, so that the effective model can be greatly simplified \cite{probing}. After mapping into a spin-1/2 model with  $\ket{g}\rightarrow\ket{\downarrow}$ and $\ket{e}\rightarrow\ket{\uparrow}$,
the many-body interacting Hamiltonian can be written as:
\begin{align}
	\hat{H}_s/\hbar=&-2\pi\delta\sum_i^{N}\hat{S}_i^z-2\pi\sum_i^N\Omega_i\hat{S}_i^x-\sum_{i\neq j}^N C_{i,j}\frac{\hat{S}_{i}^z+\hat{S}_{j}^z}{2}\nonumber\\
	-&\sum_{i\neq j}^N X_{i,j}\hat{S}_i^z\hat{S}_j^z-\sum_{i\neq j}^N J_{i,j}\hat{\bf{S}}_i\cdot\hat{\bf{S}}_j, \label{mutibodyHami}
\end{align}
where $\hat{S}^\gamma_i$ ($\gamma=x,y,z$) denotes spin operator for $i$th external state, $\delta=\omega_L-\omega_0$ is the laser detuning from atomic resonance, and $\Omega_i$ is the Rabi frequency. The coefficients of the effective spin interactions are explicitly expressed as
\begin{align}
&J_{i,j}=a_{eg}^{-}G^S_{i,j}+b_{eg}^3 G^{P}_{i,j}\nonumber\\
&C_{i,j}=(b_{ee}^3-b_{gg}^3) G^{P}_{i,j}\label{JCX}\\
&X_{i,j}=(b_{ee}^3-2 b_{eg}^3+b_{gg}^3) G^{P}_{i,j}\nonumber
\end{align}
where $G^S_{i,j}$, $G^P_{i,j}$ describes the interaction strength arising from s-wave and p-wave respectively. The derivation of the model in detail can be found in Appendix. \ref{AP_a}.
Although the Hamiltonian $\hat{H}_s$ only keeps the spin degree of freedom, the total Hilbert space is still extremely large. One solution is the collective approximation \cite{aquantum}, which takes the average of all the coefficients to replace mode dependent ones. In this approximation, the total spin is confined in the Dicke space with total spin $S=N/2$, see Appendix \ref{AP_b} for more information about collective approximation. This method is applicable when the fluctuation of all coefficients are small. However, it is no longer valid at higher temperature, because it is necessary to take the multi-bands due to the stronger thermal fluctuations. The strength of the thermal fluctuation can also be reflected by the variation of the Rabi frequency $\Delta\Omega/\overline{\Omega}$. Because the external states of the atoms are sampled according to the Boltzmann distribution, the mean value Rabi frequency can be directly calculated as $\overline{\Omega}=\sum_i^N\Omega_i/N$ and so is the standard deviation $\Delta\Omega$. As demonstrated in Fig.\ref{fig1}, $\Delta\Omega/\overline{\Omega}$ rapidly increases while the longitudinal temperature becomes higher, and it is larger than $\approx0.3$ at $T_r\gtrsim3\mu$K. To include the influence of the higher band at higher temperature, we develop the MBSED method.

\section{MBSED Method}\label{sec3}
\begin{figure}[t]
	\includegraphics[width=0.45\textwidth]{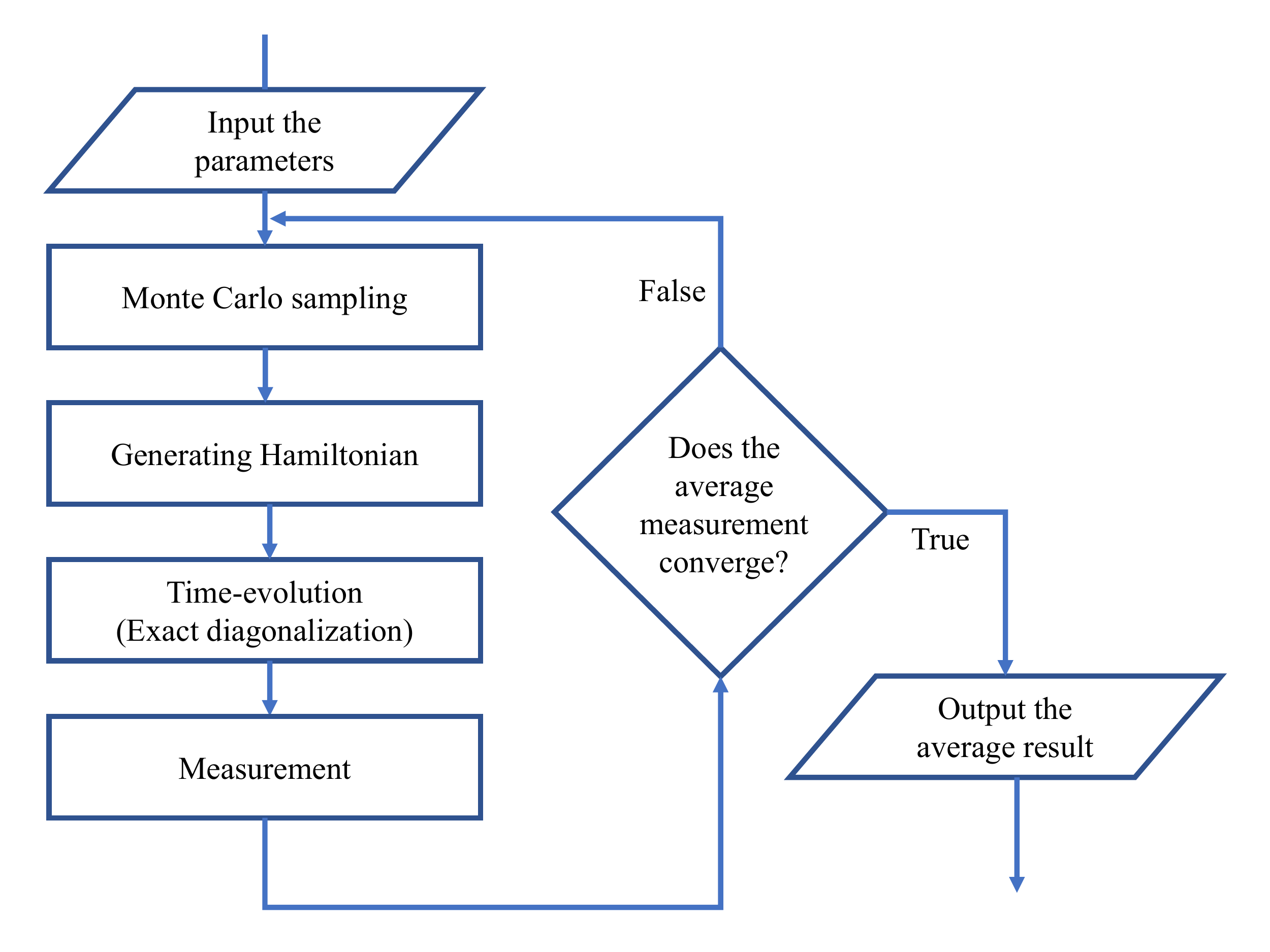}
	\caption{\label{fig2} The flow chart of the MBSED algorithm.}
\end{figure}

The algorithm of the MBSED method is pretty straightforward, and the flow chart diagram is shown in Fig.\ref{fig2}. The main processes can be concisely described as follows: (1) numbers of samples are obtained according to the Boltzmann distribution; (2) because the number of particles in each site is not too much, so that time evolution of the quantum state can be calculated via the exact diagonalization. (3) the probability of the excited state can be estimated by taking the mean value of all the samples.

We first use the Monte Carlo method to generate samples that follow the Boltzmann distribution. The corresponding partition function can be represented as:
\begin{equation}
Z=\sum_{n_z,n_r}(n_r+1)\exp(-\frac{\hbar\omega_z(n_z+\frac{1}{2})}{k_B T_z}-\frac{\hbar\omega_r(n_r+1)}{k_B T_r})\label{Boltz},
\end{equation}
in which $\omega_z$ ($\omega_r$) is the longitudinal (transverse) trap frequency, $T_z$ ($T_r$) is the longitudinal (transverse) temperature, and $n_r=n_x+n_y$ labels the transverse motional state with degeneracy $n_r+1$. The number of eigenstates $N_z$ ($N_r$) in longitudinal (transverse) direction can be estimated with $U_r/\hbar\omega_z$(($\omega_r$) where $U_r$ is the depth of lattice potential energy. Typically, $N_z$ is around $\approx 5$ and $N_r$ is around $\approx1000$. However, the motional index $(n_z,n_r)$ of each atom should be guaranteed that the total energy is less than $U_r$. And then, the motional states of all the atoms are sampled according to the distribution Eq.\ref{Boltz}. Notice that, because of the degeneracy in the transverse direction, the index $(n_x,n_y)$ should be randomly selected after fixing $(n_z,n_r)$. Meanwhile, the samples with atoms located in the same motional state $(n_x,n_y,n_z)$ have to be kicked out due to Pauli exclusion principle. Such processes are extremely rare at high temperature, and less than one per thousand at 1$\mu$K, so its influence can be omitted.

\finished After the motional index $\vec{n}$ is sampled, the corresponding coefficients of Hamiltonian (Eq.\ref{mutibodyHami}) on the basis of harmonic eigenstates can be directly calculated (see Appendix A for details). Then, all the eigenstates $|\psi\rangle_m$ with eigenenergy $E_m$ can be obtained by using the exact diagonalization. Assuming the system is prepared at the initial state $|\psi\rangle_0$ (i.e. the ground state $|\psi\rangle_G=|\downarrow \downarrow ...\downarrow\rangle$), the time-dependent wave function turns out to be 
\begin{equation}
|\psi(t)\rangle=\sum_m  e^{-iE_mt/\hbar}{}_{m} \langle \psi| \psi \rangle_0|\psi\rangle _m.
\end{equation}
However, the full diagonalization requires the atom number $N$ should be $\lesssim20$, so it is better to find a way to truncate the Hilbert space. The initial state $|\psi\rangle_G$ stays at the subspace with total spin $S=\frac{N}{2}$, but evolves into other subspaces due to the interaction and inhomogeneous excitations. The good thing is that the variation of $\langle S \rangle$ is very slow, so that we can project the Hamiltonian to the subspaces with total spin equal to $\{\frac{N}{2}, \frac{N}{2}-1,...,\frac{N}{2}-m\}$ when the evolution time is not so long. The selection of the truncation number $m$ depends on parameters of OLC such as temperatures and misalignment angle. 

The physical observables can be numerically calculated by taking the mean value of all the samples, such as the probability of the excited state. If the deviation of the mean value is converged to a certain precision, we output the final result, otherwise, the sampling will continue.

\section{Numerical simulation}\label{sec4}
The MBSED method is very useful to numerically simulate the spectroscopy with high precision in an OLC beyond the collective regime. To demonstrate its efficiency, the densities of both Ramsey spectroscopy and Rabi spectroscopy are estimated. First, the values of scattering length $b_{ee}$ and $b_{eg}$ should be fitted out by using the MBSED method. Based on the experimental data from Ye's group\cite{aquantum}, the fitting scattering lengths are $b_{ee}=150.19a_{B}$ and $b_{eg}=192.34a_{B}$ ($a_B$ is the Bohr radius). More details related to the fitting process are shown in Appendix \ref{AP_c}. Notice that, the scatting lengths we obtained are strongly different from Ref. \cite{aquantum} (different sign), and we think it may result from the different derivation of the effective model. Therefore, we include the derivation in detail in the supplementary material \cite{sup}, although it has no effect on our conclusion.
\begin{figure}[t]
	\includegraphics[width=0.38\textheight]{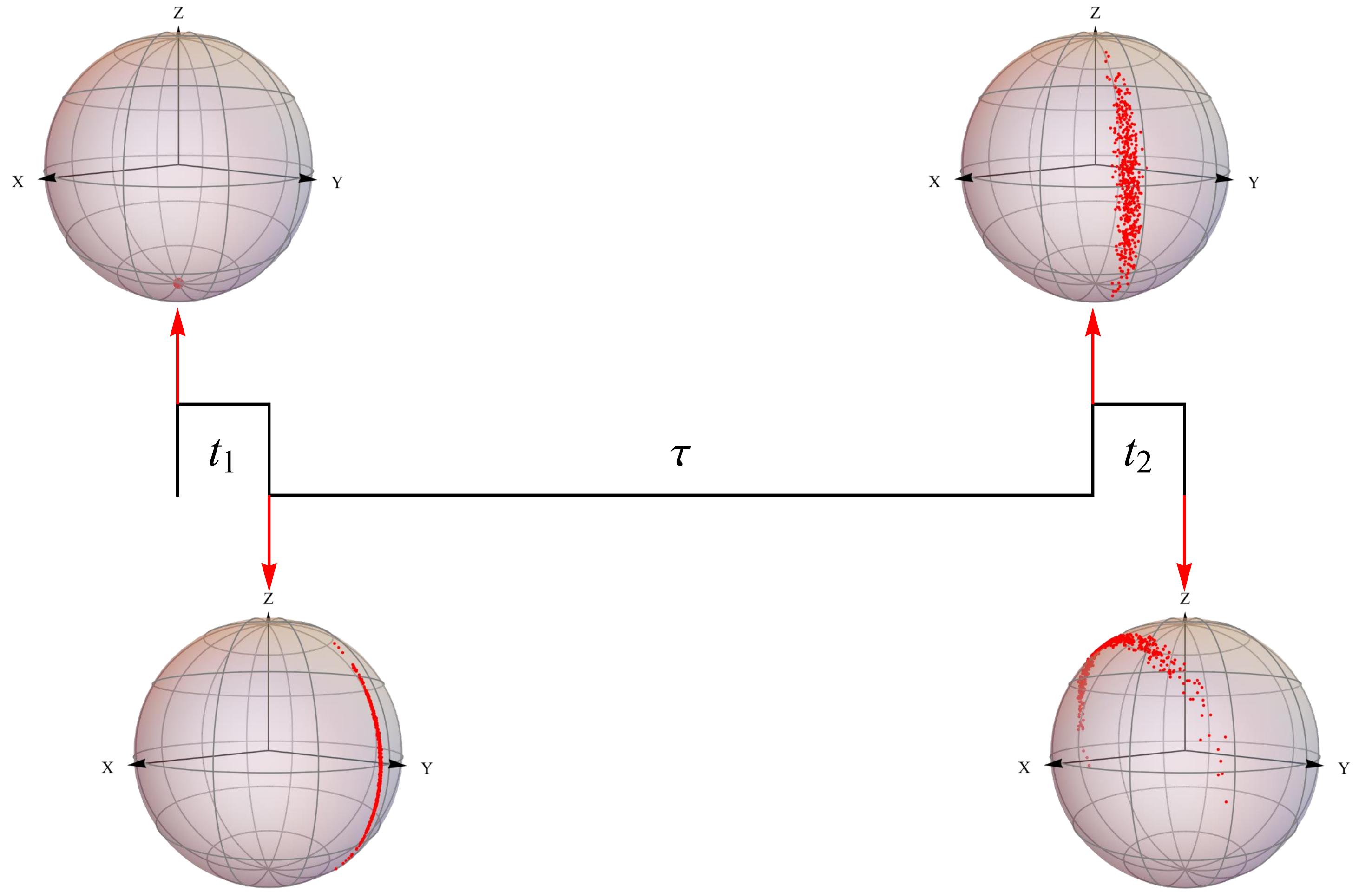}
	\caption{\label{Ramsey_Intro} A schematic diagram for Ramsey spectroscopy process of an ensemble of atoms. Here, each dot represents quantum state of one atom on the Bloch sphere. }
\end{figure} 

\subsection{Density shift in Ramsey spectroscopy}
Ramsey spectroscopy has become a well-developed tool in time measurement. Because of the laser frequency noise and excitation inhomogeneous, the atom-laser coherence time could not surpass a few seconds\cite{Half}. In comparison, the coherence time of Ramsey spectroscopy can be extended to a half-minute scale\cite{redshift}, because the atom-atom coherence is used for frequency measurement instead of atom-laser. Thus, the precision of OLC can be greatly enhanced.

In the Ramsey process, the atoms are initially prepared in the ground state. A probing pulse with detuning $\delta$ is added during the pulse time $t_1$. Then, the atoms freely evolve for a dark time $\tau$ (without probing laser).  Finally, a probing laser with the same detuning will be used for duration $t_2$. Usually, the Rabi frequency is much greater than the interaction strength in an experimental setup. Thus, the influence of inter-atomic collisions can only be taken into account in the dark time. Fig. \ref{Ramsey_Intro} shows a Ramsey process of an ensemble of atoms in the presentation of the Bloch sphere, from which one can see the inhomogeneous rotation speed to $z$ axis during the dark time due to interaction.  Because of this, the effective detuning is no longer the same for each atom, which is the origin of the density shift. 
\begin{figure}[t]
	\includegraphics[width=0.35\textheight]{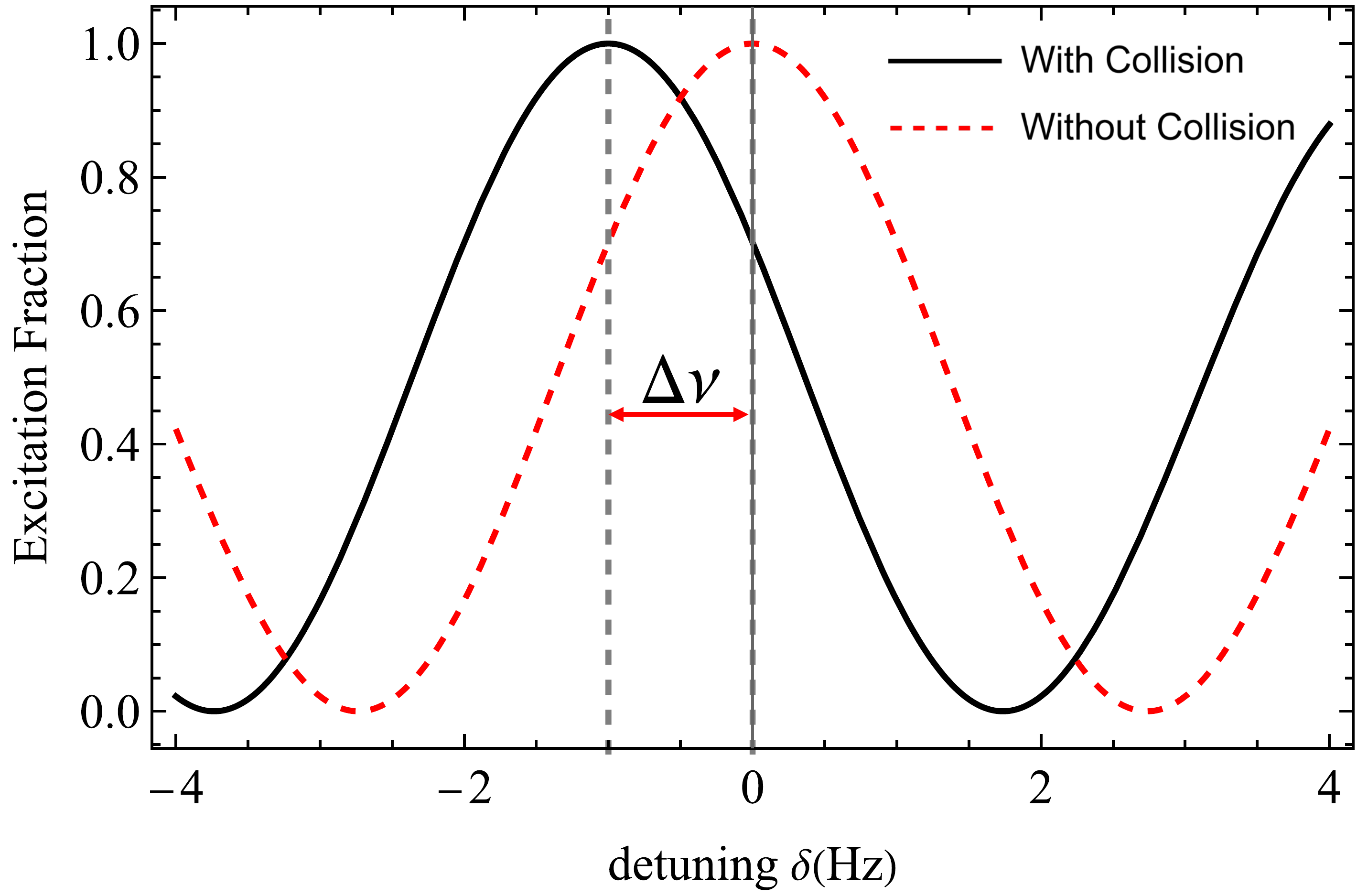}
	\caption{\label{explain} The schematic Ramsey spectrum with and without atomic collision.}
\end{figure} 

As demonstrated in Fig. \ref{explain}, the peak of the excitation fraction in the Ramsey spectroscopy locates at the zero detuning if the atomic collision is not considered. However, in the real experiment, the collision can result in the deviation of peak $\delta=2\pi\Delta\nu$ which is equal to the density shift. Under the collective approximation, the density shift could be analytically calculated  (see Appendix. \ref{AP_d}), and we will compare it with the numerical result calculated by the MBSED method. 

\begin{figure}[t]
	\includegraphics[width=0.35\textheight]{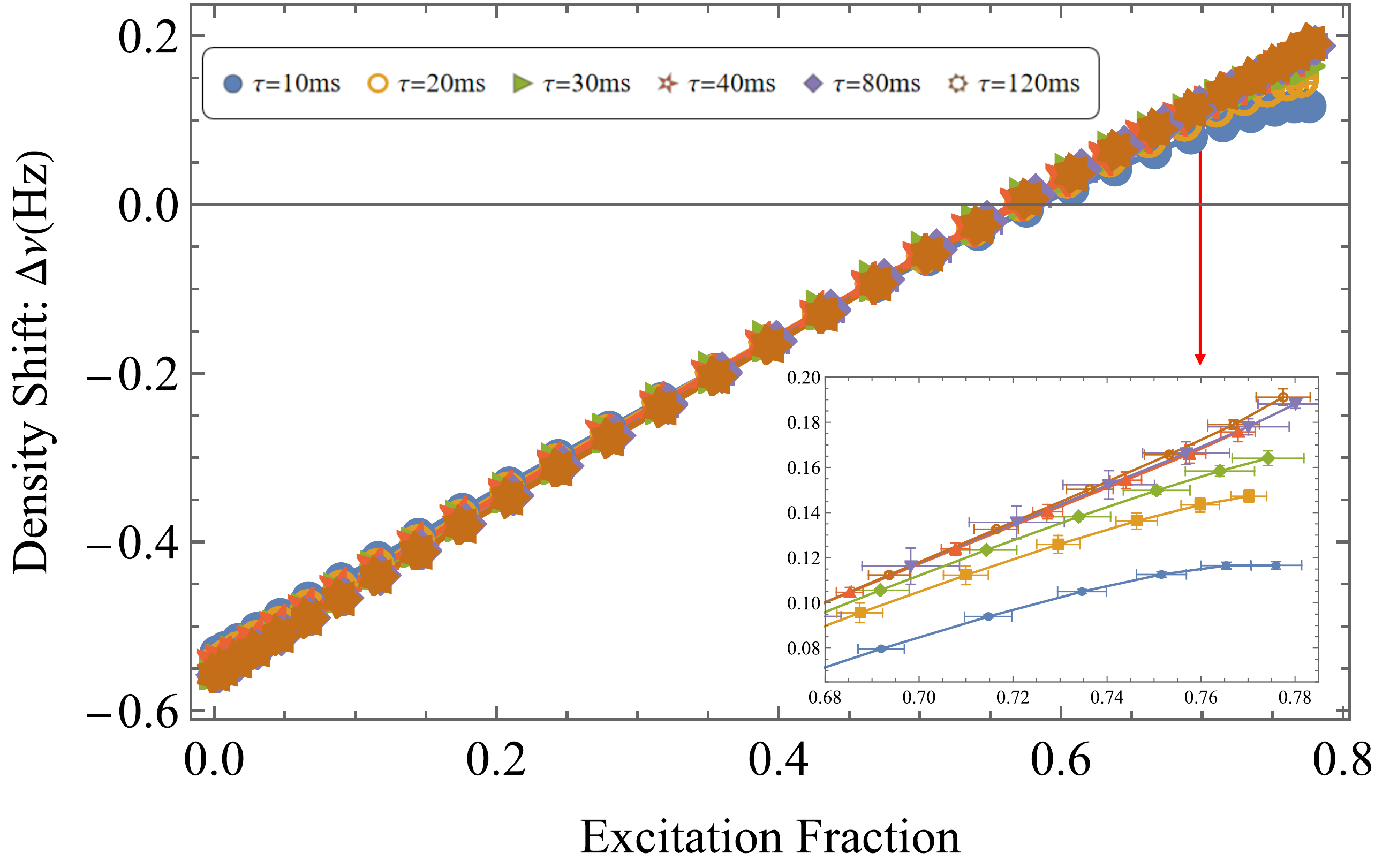}
	\caption{\label{Ramsey_tau}The density shifts of Ramsey spectroscopy at different dark time duration $\tau$. The temperature is set to be $T_z=T_r=3\mu$K and the atom number is 5. The inset shows the smaller scale.}
\end{figure} 
The time evolution of the pulse time can be strongly simplified. Because the interaction is neglectable in comparison with the remaining terms, each atom in the motional state is governed by local $2\times2$ matrix. On the other hand, during the dark time, there is no atom-light interaction. Therefore, the Hamiltonian keeps the $U(1)$ symmetry with conserved quantities $M^z=\sum_i^{N}\hat{S}_i^z$, thus could be block-diagonalized. In addition, the detuning $\delta$ will not change the eigenvector $V$ of dark time Hamiltonian, and only change the eigenvalue with a detuning related term $\hbar\delta M^z$. Considering the Ramsey spectroscopy is obtained by changing the detuning, the diagonalization can be performed just once time. In the following simulation, we set $\nu_z=66$kHz, $\nu_r=250$Hz, the misaligned angle between lattice and probing laser $\Delta\theta=10$mrad, which is a set of typical system parameters in an OLC at NSTC \cite{wang01,wang02,wang03,wang04}. The bare Rabi frequency $\Omega_0$ is set to be 500Hz, and has less effect on the density shift since it is much larger than the collision coefficients. In order to figure out the relation between density shift and the population of the excitations, we vary the excitation fraction at the end of the first pulse by tuning $t_1$ at fixed $t_2=\frac{\pi}{2\overline{\Omega}}$.

\begin{figure}[b]
	\includegraphics[width=0.35\textheight]{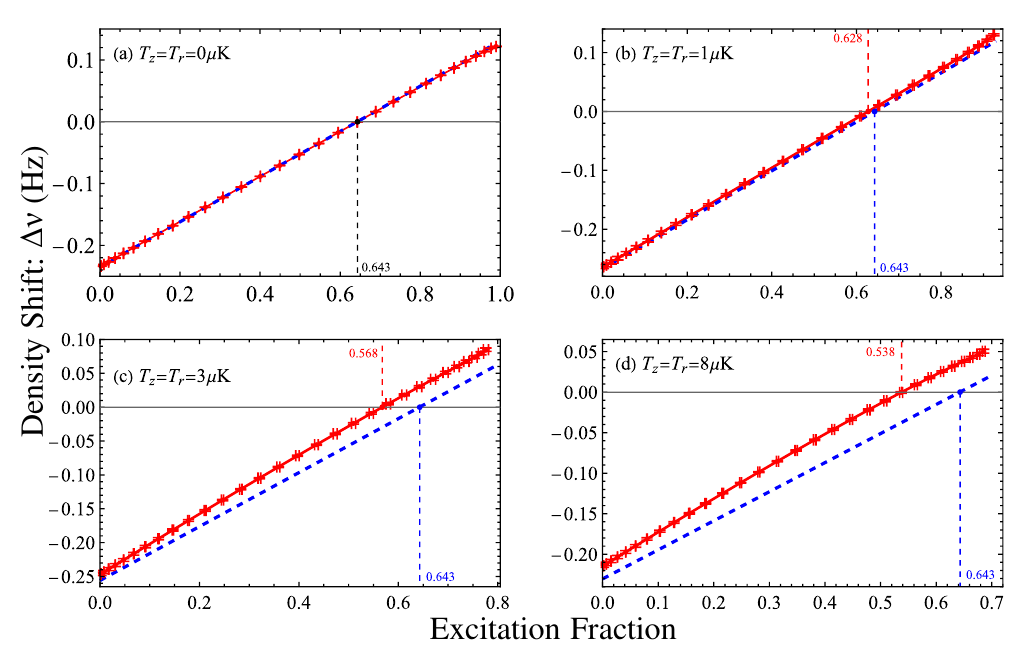}
	\caption{\label{Ramsey_ED_VS_CA_T}Comparison of density shift between numerical simulation (red solid line) and collective approximation (blue dashed line) at different temperatures with the atom number equal to 5. The vertical dash lines mark the excitation fraction with the zero density shift.}
\end{figure}  
First, the playing role of the dark time is checked. In the collective approximation, it has no influence on the dark time. However, we can clearly find different $\tau$ can cause different density shift from Fig. \ref{Ramsey_tau} at higher temperature, especially when the dark time is short. Thus, to eliminate the influence of the dark time, we set  $\tau=120ms$ around which the deviation of the density shift becomes small.

\begin{figure}[t]
	\includegraphics[width=0.35\textheight]{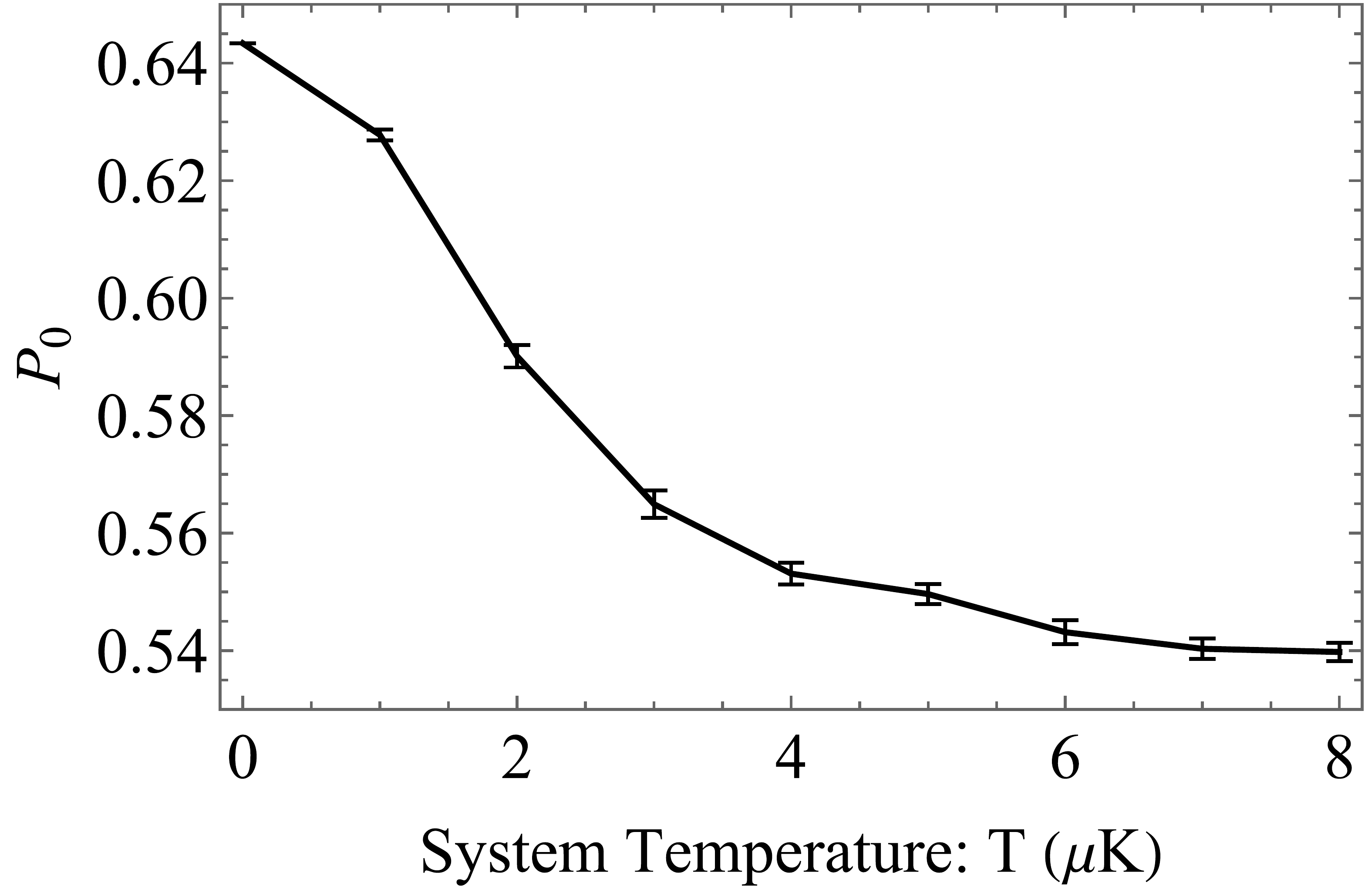}
	\caption{\label{Ramsey_T} The excitation fraction for zero density shift at different temperatures. The system parameters are the same as Fig.\ref{Ramsey_ED_VS_CA_T}.}
\end{figure}

Then, as mentioned before, the temperature can result in the invalid of the collective approximation, so it is worth verifying it. In Fig. \ref{Ramsey_ED_VS_CA_T}, we compare the results of both collective approximation and MBSED methods at different temperatures. It is obvious that there is a large deviation between them while $T_z=T_r>3\mu K$. From Fig. \ref{Ramsey_ED_VS_CA_T}, we can extract the excitation fraction $P_0$ where the corresponding density shift equals zero. The collective approximation gives the temperature-independent value $P_0=0.643$. In contrast, as shown in Fig. \ref{Ramsey_T}, the simulation from MBSED method presents $P_0$ decreases while temperature increases. Therefore, the MBSED method can remedy the drawback of the collective approximation and provide more feasible experimental conditions at a higher temperature. 

\begin{figure}[b]
	\includegraphics[width=0.3\textheight]{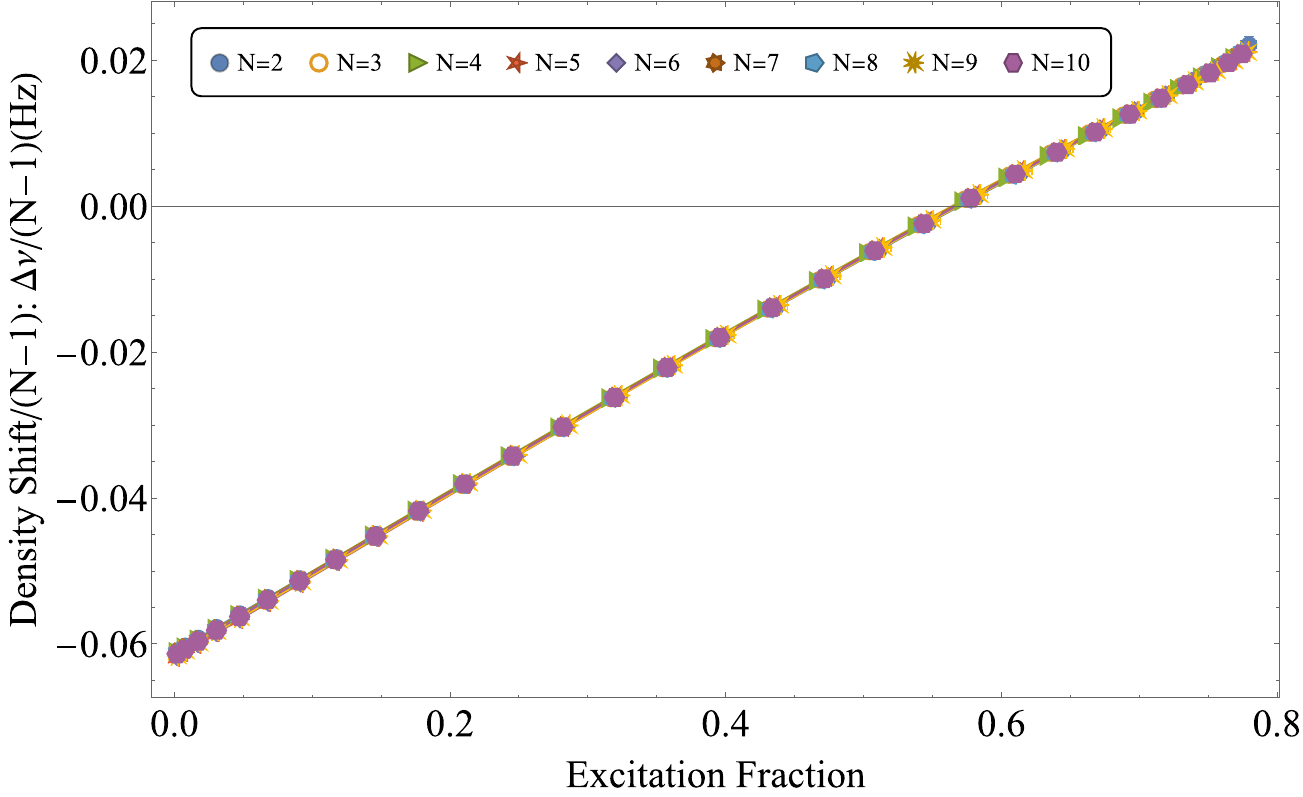}
	\caption{\label{ESRN} Rescaled density shift in Ramsey spectroscopy with different numbers of atoms. The temperature is set to be $T_z=T_r=3\mu K$. Inset: the density shift (red dot) for different atom number at excitation fraction $P_e=0.2$, and the solid line is the linear fitting result. }
\end{figure} 
At last, we want to check the linear relation between density shift and atom number predicted by collective approximation, because this relation is usually used to estimate the density shift in the experiment \cite{probing}. Meanwhile, we also use this relation to fitting the system parameters (see Appendix. \ref{AP_c}). In Fig. \ref{ESRN}, after being divided by $N-1$, the curves of density shift of different atom numbers almost overlap with each other, so that the excitation fraction with zero density shift is almost unchanged with atom number.  To be more explicit, we performed a linear fit of the density shift at excitation fraction $P_e=0.2$ for different atom numbers. As shown in the inset of Fig. \ref{ESRN}, the linearity of the relation is very good. Considering the corresponding temperature is $T_r=T_z=3\mu K$, it means the linear property of density shift to atom number in Ramsey spectroscopy remains for high temperature.

\begin{figure}[t]
 	\includegraphics[width=0.5\textwidth]{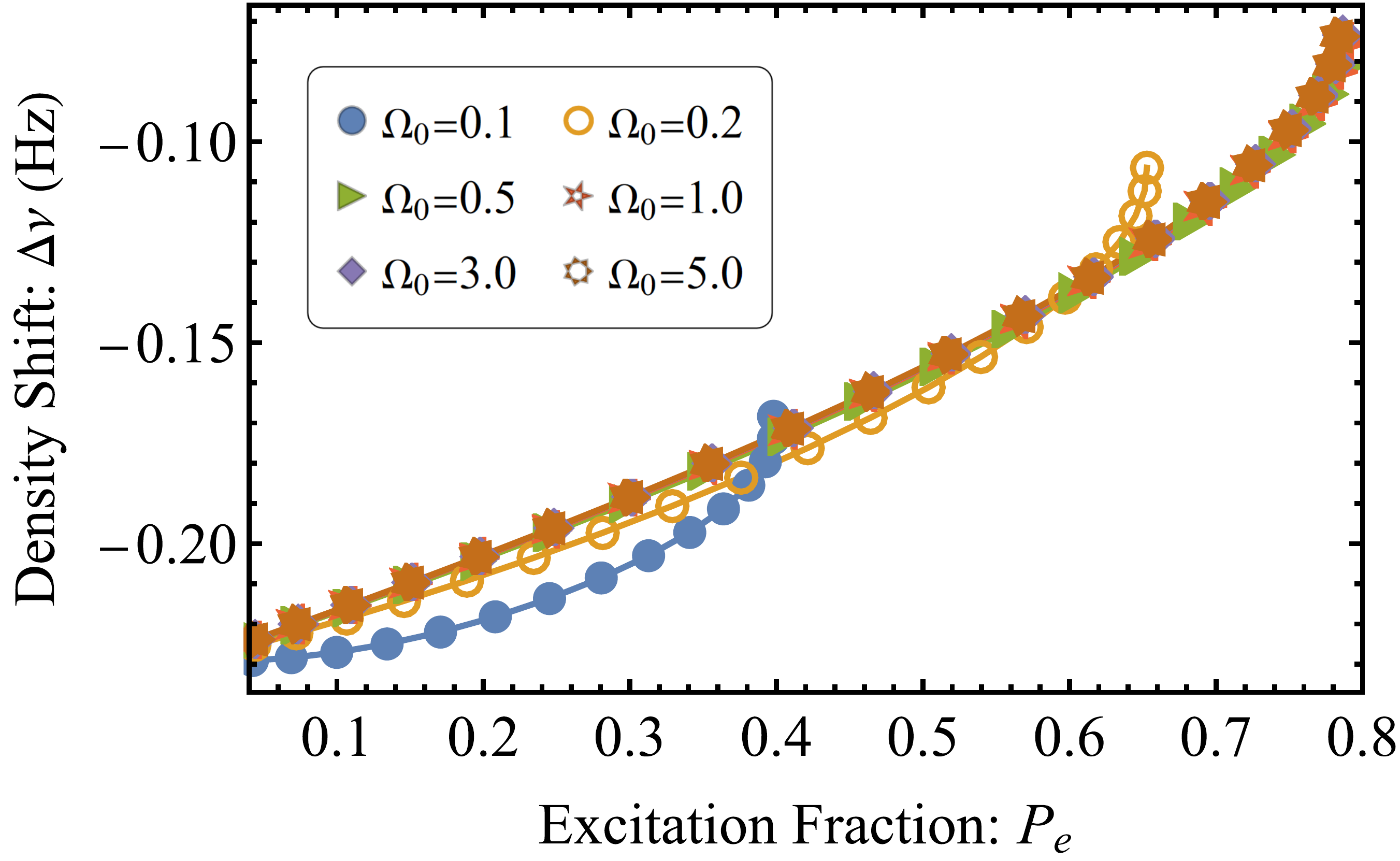}
 	\caption{\label{Rabi_G} Density shift of Rabi spectroscopy at different Rabi frequencies. The temperature is set to be $T_z=T_r=3\mu$K and the atom number is 5.}
\end{figure}
\subsection{Density shift in Rabi spectroscopy}
In Rabi spectroscopy, the atoms are also initially prepared in the ground state and then excited by applying a probe laser with detuning $\delta$ with duration time $t$. In comparison with Ramsey spectroscopy, the Rabi frequency is around several Hz, so the collision during pulse time can not be ignored. Meanwhile, because the total spin is not conserved, much more computational effort is required than in the Ramsey case. Thus, we only consider the density shift within a $\pi$ pulse. The density shift $2\pi\Delta \nu$ is defined as the detuning of the maximum excitation point in Rabi spectroscopy.

\begin{figure}[b]
	\includegraphics[width=0.45\textwidth]{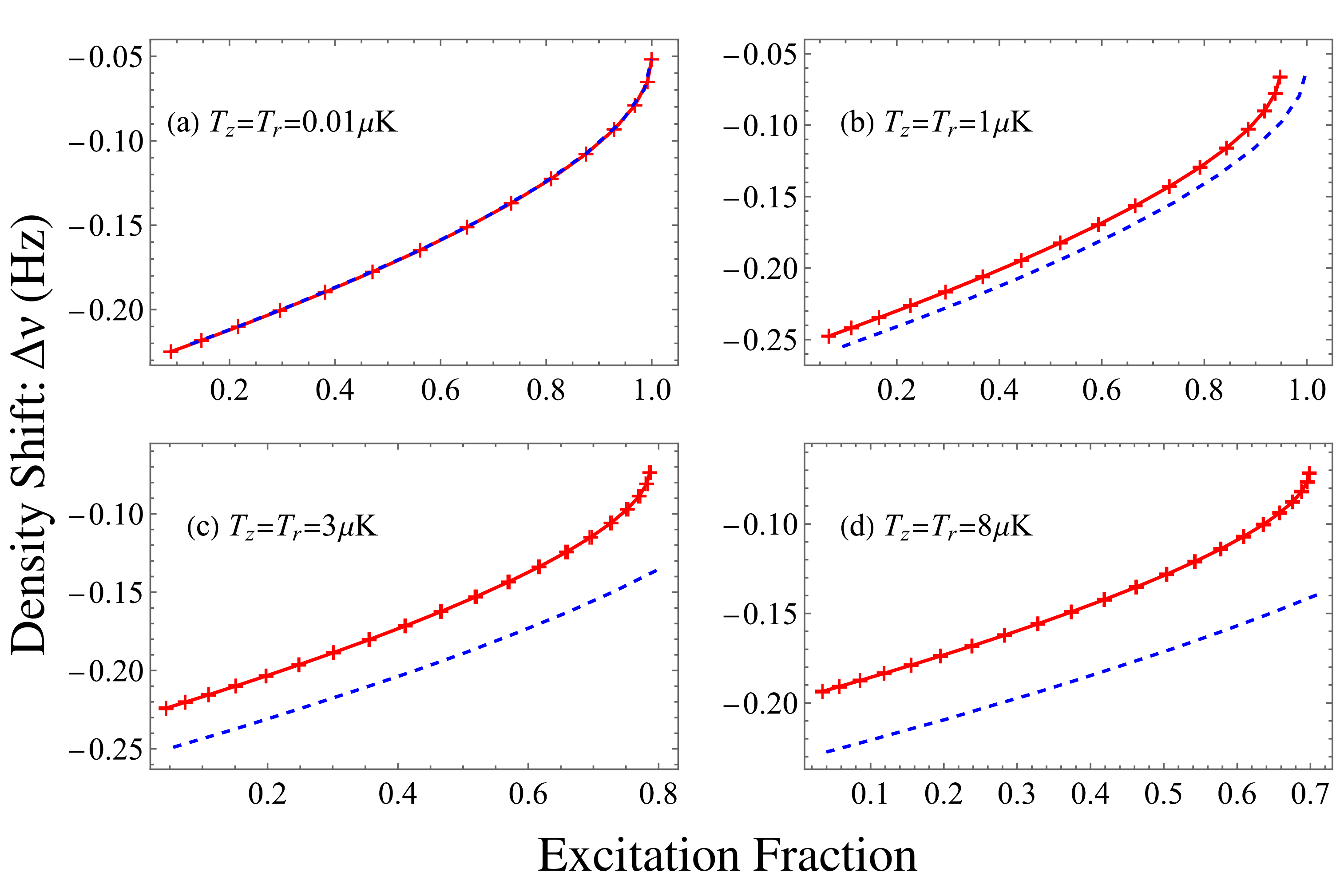}
	\caption{\label{Rabi_ED_VS_MF}Comparison of density shift between numerical simulation (red solid line) and collective approximation (blue dashed line) at different temperatures with the atom number equal to 5.}
\end{figure}
First, the relation between density shift and Rabi frequency is checked. From Fig. \ref{Rabi_G}, we can find the density heavily depends on the magnitude of the Rabi frequency. However, when the Rabi frequency is a magnitude larger than the interatomic collision energy scale which is around 0.06 Hz, the density shift converged to a certain value. Thus, we set the bare Rabi frequency $\Omega_0=5$ Hz in the following discussion. Furthermore, we can notice that the relation between density shift and excitation fraction is no longer linear, which is different from Ramsey spectroscopy.

Then, we want to study the influence of temperature. The collective approximation can also be used in Rabi spectroscopy, but the time-evolution can not be solved analytically. Therefore, we numerically calculated the density shift under collective approximation and compared them with the MBSED method. Fig. \ref{Rabi_ED_VS_MF} shows the density shift of both methods at different temperatures. Same as the Ramsey case, it indicates that the collective approximation is only suitable for low temperatures.  At higher temperatures, it will overestimate the density shift.

\begin{figure}[t]
	\includegraphics[width=0.45\textwidth]{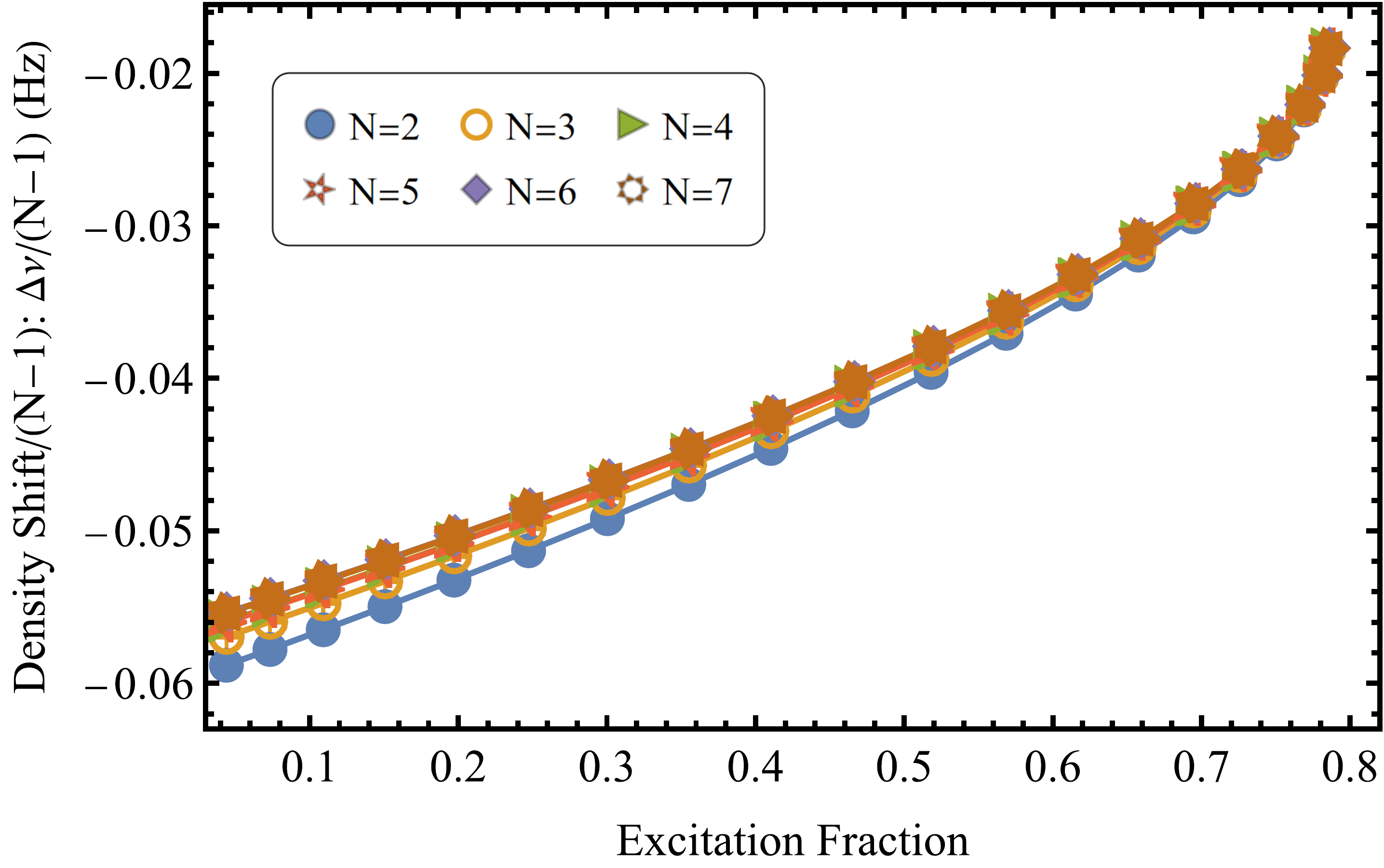}
	\caption{\label{Rabi_N} Rescaled density shift in Rabi spectroscopy with different numbers of atoms. The temperature is set to be $T_z=T_r=3\mu K$. }
\end{figure}

At last, we also check if the density shift has a linear relation with atom number. From the Fig. \ref{Rabi_N}, we can find the rescaled density shift with different numbers of atoms only overlap at high excitation fraction region. If the excitation is lower, the linear relation is not valid. 

\section{CONCLUSION and OUTLOOK}\label{sec5}
In this paper, we extended the effective spin Hamiltonian which describes the atomic collision of OLC from single band to the multi-band case, so that the OLC system at higher temperatures can be described. To deal with the extended Hamiltonian, we developed a numerical algorithm-MBSED, which combines the Monte-Carlo method to sample the distribution of atoms and exact diagonalization to simulate the time evolving of the OLC system. After implementing the MBSED method on both Ramsey and Rabi spectroscopies, we found the collective approximation method is only valid for the system at low temperatures. For the Ramsey spectroscopy, we found that the linear relation between density shift and atom number still holds, and the special excitation fraction where density shift equals zero decreases with increasing temperature. For the Rabi case, we found that the relation between density shift and atom number is no longer linear. Meanwhile, there is no special excitation fraction where the density shift equals zero, although the density shift becomes smaller when the excitation fraction is higher. Thus we suggest using the highest excitation of the Rabi spectrum for precision measurement.

The quantum many-spin model has been extensively studied in the field of condensed matter physics. In order to solve such quantum many-body problems, numerous numerical algorithms are developed, including exact diagonalization, quantum Monte Carlo, matrix product states, and so on. However, in the field of OLC, there is no trial that borrows these algorithms into solving the quantum many-body Hamiltonian. Thus, our work provides a bridge between OLC and quantum many-body computation. Then, a lot of questions could be asked in future work. One direction is checking how average entanglement entropy grows with the atom number to decide whether the model could deal with the matrix product state or not.  Another direction would be further considered in-elastic collision between atoms which has been ignored in this paper.

\section{ACKNOWLEDGMENTS}
We wish to thank A. M. Rey for the discussions.  This work is supported by the National Science Foundation of China under Grants  No. 12274045 and China Postdoctoral Science Foundation Funded Project No. 2020M673118. X.-F. Z. acknowledges funding from the National Science Foundation of China under Grants  No. 12274046, No. 11874094 and No.12147102, Fundamental Research Funds for the Central Universities Grant No. 2021CDJZYJH-003.  T. Wang acknowledges funding supported by the Program of State Key Laboratory of Quantum Optics and Quantum Optics Devices(No:KF202211).

\newpage
\appendix
\section{Effective Hamiltonian}\label{AP_a}
\begin{figure}[t]
	\includegraphics[width=0.45\textwidth]{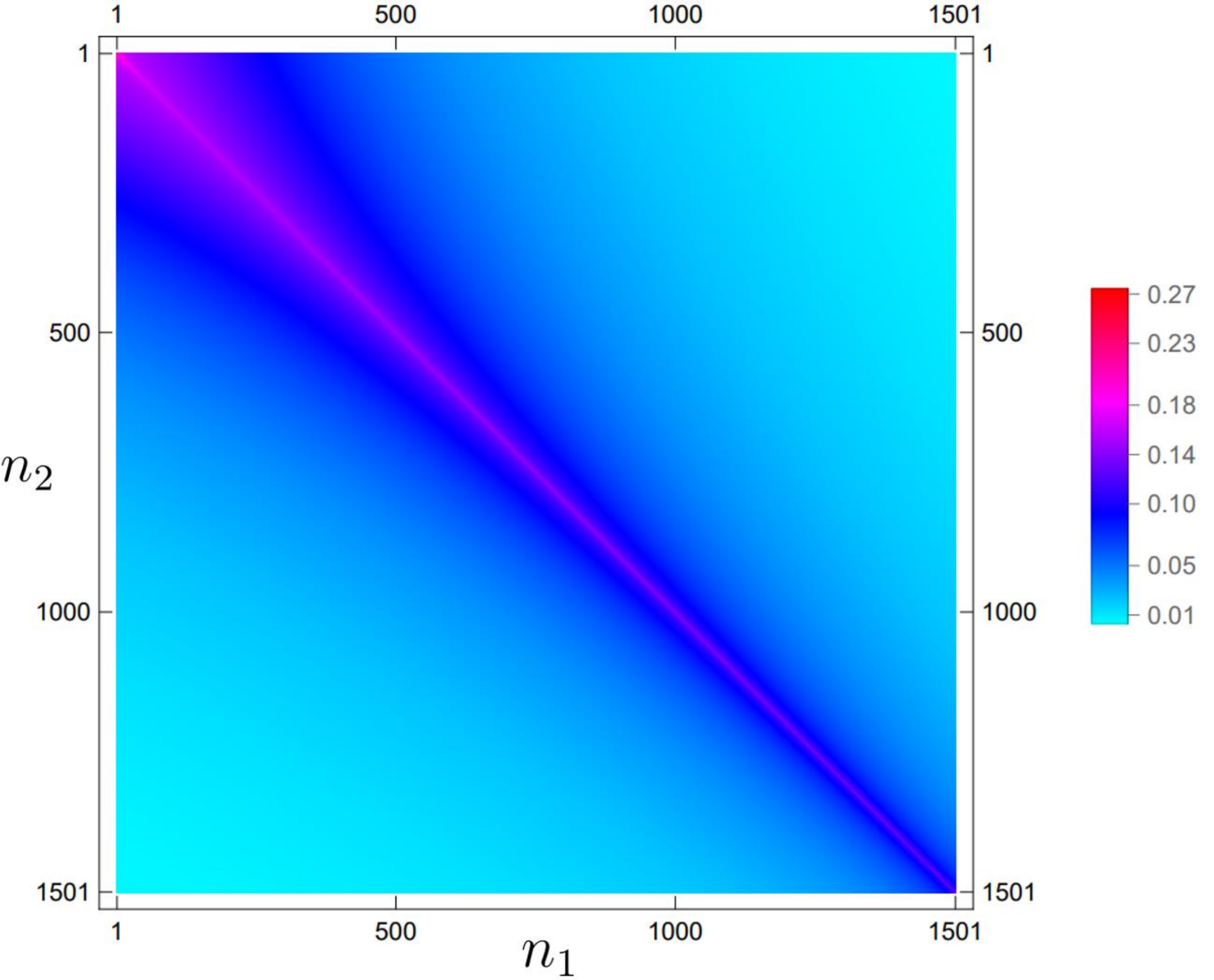}
	\caption{\label{s-function}The mode dependence of $s(n_1,n_2,n_1,n_2)$}
\end{figure}

\begin{figure}[t]
	\includegraphics[width=0.45\textwidth]{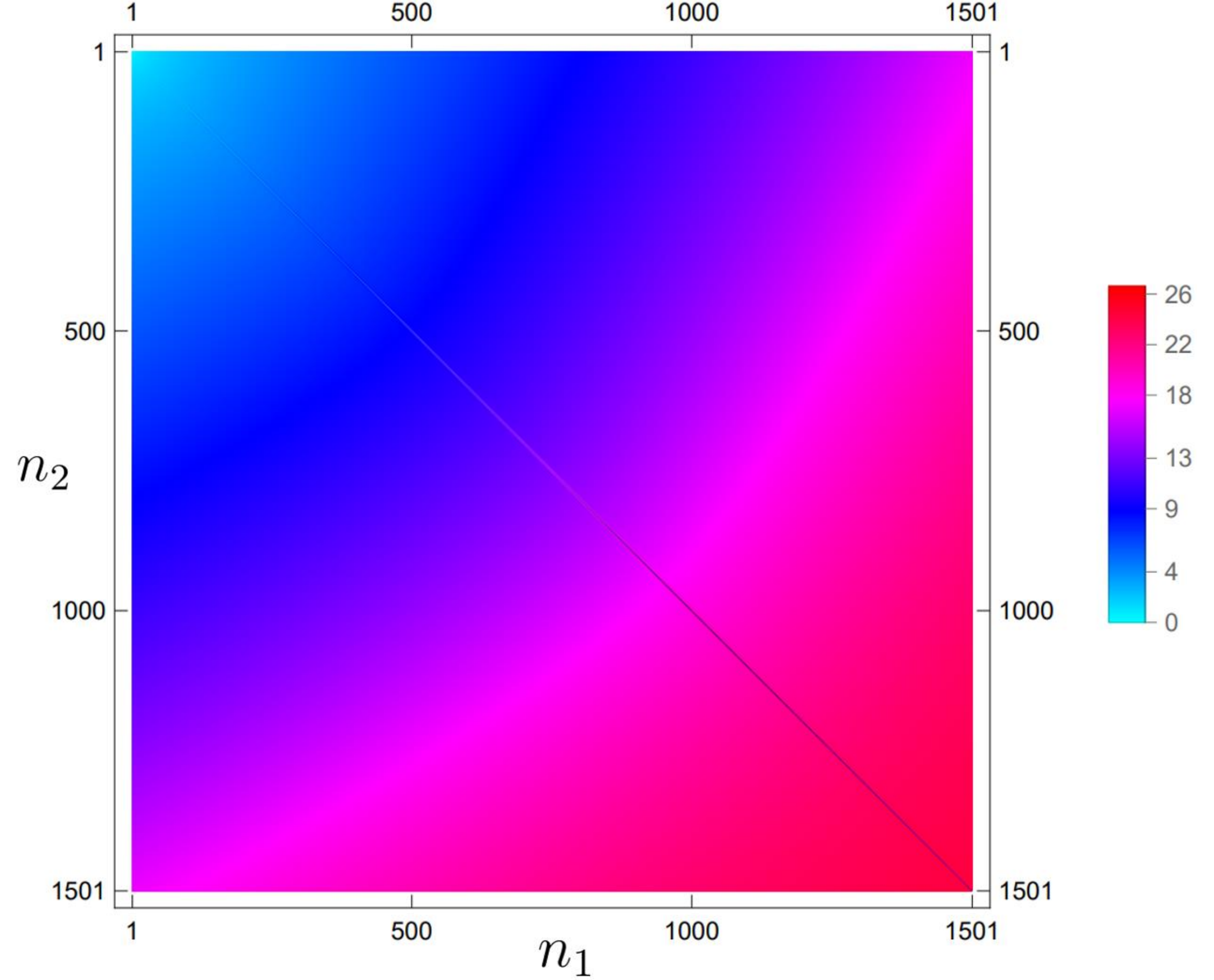}
	\caption{\label{p-function}The mode dependence of $p(n_1,n_2,n_1,n_2)$}
\end{figure}

After expanding the Hamiltonian (Eq.\ref{OriginalHamiltonian}) on the non-interaction harmonic basis (Eq.\ref{basis}), we get Hamiltonian:
\begin{align*}
	\hat{H}_h=&\hat{H}_c+\frac{1}{2}\hbar\omega_0\sum_{i}^N(\hat{c}_{e,\vec{n}_i}^{\dagger}\hat{c}_{e,\vec{n}_i}-\hat{c}_{g,\vec{n}_i}^{\dagger}\hat{c}_{g,\vec{n}_i})\\
	-&\frac{1}{2}\hbar\sum_{i}^{N}\Omega_{\vec{n}_i}(e^{-i \omega_L t}\hat{c}_{e,\vec{n}_i}^{\dagger}\hat{c}_{g,\vec{n}_i}+e^{i \omega_L t}\hat{c}_{g,\vec{n}_i}^{\dagger}\hat{c}_{e,\vec{n}_i})
\end{align*}
Here, $\Omega_{\vec{n}}$ is the Rabi frequency at motional state $\vec{n}$ considering the Doppler effect \cite{dingbiao} and $\hat{H}_c$ is the collision Hamiltonian given later. Let us define:
\begin{align*}
&s(n_1,n_2,n_3,n_4)=
\\
&\int\frac{H_{n_1}(\xi)H_{n_2}(\xi)H_{n_3}(\xi)H_{n_4}(\xi)}{\pi\sqrt{2^{n_1+n_2+n_3+n_4}n_1!n_2!n_3!n_4!}}e^{-2\xi^2}d\xi\\
&p(n_1,n_2,n_3,n_4)=
\\
&\int\frac{(\frac{dH_{n_1}}{d\xi}H_{n_2}-H_{n_1}\frac{dH_{n_2}}{d\xi})(\frac{dH_{n_3}}{d\xi}H_{n_4}-H_{n_3}\frac{dH_{n_4}}{d\xi})}{\pi\sqrt{2^{n_1+n_2+n_3+n_4}n_1!n_2!n_3!n_4!}}e^{-2\xi^2}d\xi
\end{align*}
where $H_n(\xi)$ are Hermite polynomials.
The coefficients $s(n_1,n_2,n_3,n_4)$ could be recursively calculated:
\begin{align*}
	s(n_1+2,n_2,n_3,n_4)=&\frac{1}{2}\sqrt{\frac{n_2}{n_1+2}}s(n_1+1,n_2-1,n_3,n_4)\\
	+&\frac{1}{2}\sqrt{\frac{n_3}{n_1+2}}s(n_1+1,n_2,n_3-1,n_4)\\
	+&\frac{1}{2}\sqrt{\frac{n_4}{n_1+2}}s(n_1+1,n_2,n_3,n_4-1)\\
	-&\frac{1}{2}\sqrt{\frac{n_1+1}{n_1+2}}s(n_1,n_2,n_3,n_4),
\end{align*}
and the $p$ could be calculated from $s$: 
\begin{align*}
	p(n_1,n_2,n_3,n_4)=&2\sqrt{n_1n_3}s(n_1-1,n_2,n_3-1,n_4)\\
	-&2\sqrt{n_2n_3}s(n_1,n_2-1,n_3-1,n_4)\\
	-&2\sqrt{n_1n_4}s(n_1-1,n_2,n_3,n_4-1)\\
	-&2\sqrt{n_2n_4}s(n_1,n_2-1,n_3,n_4-1).
\end{align*}
Because the motional degrees of freedom are effectively frozen due to the energy conservation \cite{probing}, only terms satisfying $n_1=n_3$, $n_2=n_4$ or $n_1=n_4$, $n_2=n_3$ need to be considered. The dimension of the parameter space of $s$ and $p$ could be squeezed from four to two.  In Figure. \ref{s-function} and \ref{p-function}, we show the mode dependence of $s(n_1,n_2,n_1,n_2)$ and $p(n_1,n_2,n_1,n_2)$. Then, after defining 
\begin{align*}
	&S_{\vec{n}_i,\vec{n}_j}=\\
	&s(n_{x_i},n_{x_j},n_{x_j},n_{x_i})s(n_{y_i},n_{y_j},n_{y_j},n_{y_i})s(n_{z_i},n_{z_j},n_{z_j},n_{z_i})
\end{align*}
\begin{align*}
	&P_{\vec{n}_i,\vec{n}_j}^R=\\
	&p(n_{x_i},n_{x_j},n_{x_j},n_{x_i})s(n_{y_i},n_{y_j},n_{y_j},n_{y_i})s(n_{z_i},n_{z_j},n_{z_j},n_{z_i})+\\
	&s(n_{x_i},n_{x_j},n_{x_j},n_{x_i})p(n_{y_i},n_{y_j},n_{y_j},n_{y_i})s(n_{z_i},n_{z_j},n_{z_j},n_{z_i})
\end{align*}
\begin{align*}
	&P_{\vec{n}_i,\vec{n}_j}^Z=\\
	&s(n_{x_i},n_{x_j},n_{x_j},n_{x_i})s(n_{y_i},n_{y_j},n_{y_j},n_{y_i})p(n_{z_i},n_{z_j},n_{z_j},n_{z_i})
\end{align*}
the collision Hamiltonian could be written as:
\begin{align*}
	\hat{H}_c=&\frac{2\pi\hbar^2}{m}a_{eg}^{-}\sum_{\vec{n}_1\neq\vec{n}_2}R_r^2R_zS_{\vec{n}_1,\vec{n}_2}\times\\
	(-&\hat{c}_{e,\vec{n}_1}^{\dagger}\hat{c}_{g,\vec{n}_1}\hat{c}_{g,\vec{n}_2}^{\dagger}\hat{c}_{e,\vec{n}_2}+\hat{c}_{e,\vec{n}_1}^{\dagger}\hat{c}_{e,\vec{n}_1}\hat{c}_{g,\vec{n}_2}^{\dagger}\hat{c}_{g,\vec{n}_2}\\
	-&\hat{c}_{g,\vec{n}_1}^{\dagger}\hat{c}_{e,\vec{n}_1}\hat{c}_{e,\vec{n}_2}^{\dagger}\hat{c}_{g,\vec{n}_2}+\hat{c}_{g,\vec{n}_1}^{\dagger}\hat{c}_{g,\vec{n}_1}\hat{c}_{e,\vec{n}_2}^{\dagger}\hat{c}_{e,\vec{n}_2})\\
	-&\frac{3\pi\hbar^2}{m}\sum_{\alpha,\beta}b_{\alpha\beta}^3\sum_{\vec{n}_1\neq\vec{n}_2}(R_r^4R_zP_{\vec{n}_1,\vec{n}_2}^R+R_r^2R_z^3P_{\vec{n}_1,\vec{n}_2}^Z)\times\\
	&(\hat{c}_{\alpha,\vec{n}_1}^{\dagger}\hat{c}_{\beta,\vec{n}_2}^{\dagger}\hat{c}_{\beta,\vec{n}_2}\hat{c}_{\alpha,\vec{n}_1}-\hat{c}_{\alpha,\vec{n}_1}^{\dagger}\hat{c}_{\beta,\vec{n}_2}^{\dagger}\hat{c}_{\beta,\vec{n}_1}\hat{c}_{\alpha,\vec{n}_2})
\end{align*}
where $R_r=\sqrt{\frac{m\omega_r}{\hbar}}$, $R_z=\sqrt{\frac{m\omega_z}{\hbar}}$. By implementing following mapping $\ket{g}\rightarrow\ket{\downarrow}$, $\ket{e}\rightarrow\ket{\uparrow}$, the total Hamiltonian $\hat{H}_h$ is transformed to an effective spin Hamiltonian (Eq. \ref{mutibodyHami}) with interaction strengths in Eq.\ref{JCX} expressed as:
\begin{align*}
	&G_{i,j}^S=\frac{4\pi\hbar}{m}R_r^2R_zS_{\vec{n}_1,\vec{n}_2}\\
	&G_{i,j}^P=\frac{6\pi\hbar}{m}(R_r^4R_zP_{\vec{n}_1,\vec{n}_2}^R+R_r^2R_z^3P_{\vec{n}_1,\vec{n}_2}^Z).
\end{align*}

\section{Collective Approximation}\label{AP_b}
The interaction strengths ($G_{i,j}^S$, $G_{i,j}^P$) and Rabi frequency $\Omega_{i}$ are related to the motional degree of freedom. However, the standard deviations of them ($\Delta G_{i,j}^S$, $\Delta G_{i,j}^P$, $\Delta \Omega_{i}$) are very small in a collective regime, which requires the temperature is low and the misaligned angle is tiny. Therefore, the collective approximation can be utilized, and it stands for that these mode-dependent parameters can be replaced with their average value. Then, the Hamiltonian (Eq.\ref{mutibodyHami}) can be approximately written as:
\begin{align*}
	\hat{H}_\textrm{col}/\hbar=-2\pi\delta\hat{S}^z-2\pi\overline{\Omega}\hat{S}^x-\overline{X}\hat{S}^z\hat{S}^z-(N-1)\overline{C}\hat{S}^z
\end{align*}
where
\begin{align*}
	&\hat{S}^{\gamma=x,y,z}=\sum_{i}^N\hat{S}_i^\gamma\qquad\overline{\Omega}=\frac{\sum_{i}^N\Omega_i}{N}\\
	&\overline{X}=\frac{\sum_{i\neq j}^NX_{i,j}}{N(N-1)}\qquad\overline{C}=\frac{\sum_{i\neq j}^NC_{i,j}}{N(N-1)}.
\end{align*}
Under the collective approximation, the quantum state is restricted in the sub Hilbert space with total spin $S=\frac{N}{2}$. Therefore, the Heisenberg term in Hamiltonian (Eq.\ref{mutibodyHami}) is a constant and can be ignored.
\begin{figure}[t]
	\includegraphics[width=0.45\textwidth]{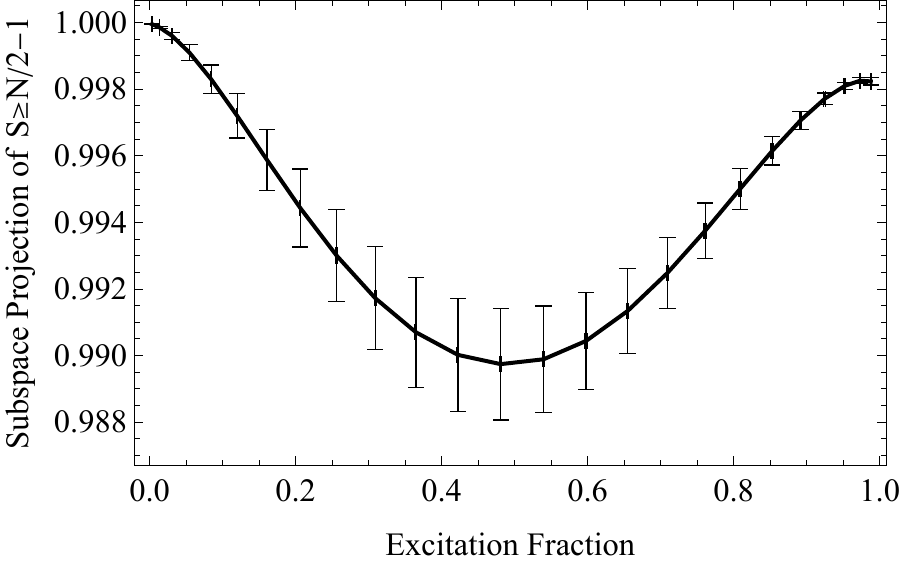}
	\caption{\label{Ramsey_S} The numerical results of projection in the subspace with total spin equal to $S=\frac{N}2$ and $S=\frac{N}2-1$ for different excitation fraction at the end of dark time in the Ramsey spectroscopy. The number of atoms is 12, and the other parameters are the same as the experiment \cite{aquantum}.}
\end{figure}

\section{Fitting of the Scattering Length}\label{AP_c}
The experiment we considered is Ref. \cite{aquantum} from Ye's group, so the parameters are set to be $\Delta\theta=5$mrad, $\nu_r=450$Hz,  $\nu_z=80$kHz, $T_r=3\mu $K, and $T_z=1.5\mu $K. In the experiment \cite{aquantum}, the average number of atoms in each site is about 20 which is still large for our method. Thus, we set the atom numbers to be 12, and obtain the density shift of 20 atoms after rescaling due to the linear relation in high density. Considering the low-temperature and small misaligned angle $\Delta\theta$, the truncation of Hilbert space could be used. As demonstrated in Fig.\ref{Ramsey_S}, only two subspace need to be taken into account.

\begin{figure}[h]
	\includegraphics[width=0.45\textwidth]{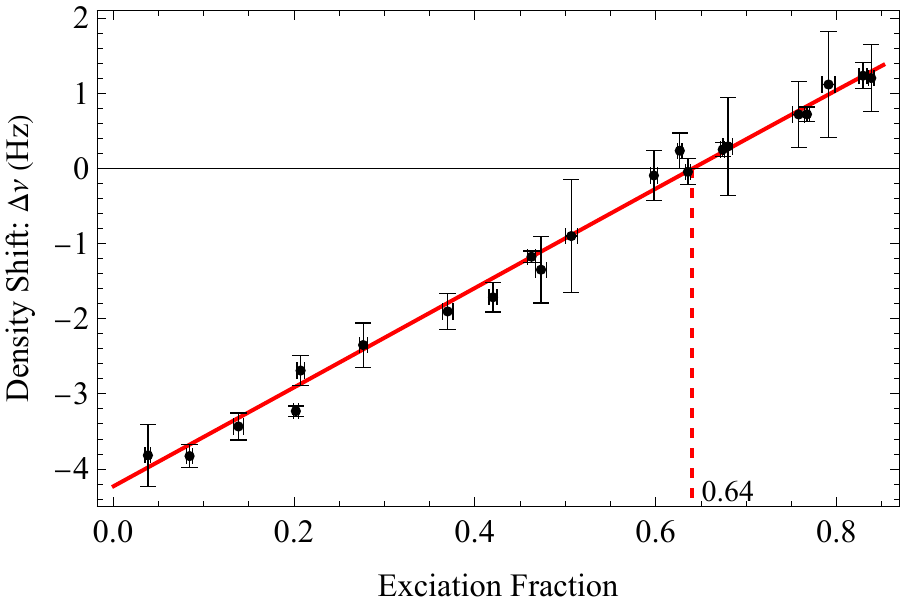}
	\caption{\label{Fit} The experimental results in Ref.\cite{aquantum} (black dots) and the fitting numerical results (red solid line) of the density shift at different excitation fraction. The fitted scattering lengths are $b_{eg}=192.34a_B$ and $b_{ee}=150.19a_B$.}
\end{figure}

After calculating the density shift of 12 atoms, we can obtain the density shift of 20 atoms by multiply a factor of $\frac{19}{11}$. As shown in Fig. \ref{Fit}, the experimental data and the fitting data match well, while the fitted scattering lengths are $b_{eg}=192.34a_B$ and $b_{ee}=150.19a_B$.

\section{Analytic Calculation of Density Shift under the Collective Approximation}\label{AP_d}
The density shift in the Ramsey spectroscopy could be approximately analytically calculated. According to the section \ref{sec4}, the time-evolution operator in Ramsey process could be written as
$$
\hat{U}(t_1,\tau,t_2)=e^{-i t_2 \hat{H}_{p}/\hbar}e^{-i \tau \hat{H}_{d}/\hbar}e^{-i t_1 \hat{H}_{p}/\hbar},
$$
where the Hamiltonian during the pulse time is
$$
\hat{H}_{p}/\hbar=-2\pi\overline{\Omega}\hat{S}^x,
$$
and the Hamiltonian during the dark time can be expressed as
$$
\hat{H}_{d}/\hbar=-2\pi\delta\hat{S}^z-\overline{X}\hat{S}^z\hat{S}^z-(N-1)\overline{C}\hat{S}^z.
$$
Here, approximately, $\hat{H}_{p}$ does not consider atomic collision and detuning, and $\hat{H}_{d}$ does not consider atom-light interaction. We use $\ket{t_1,\tau,t_2}=\hat{U}(t_1,\tau,t_2)\ket{0}$ to denote the quantum state during Ramsey process, and the measurement at the end of the dark time should be
\begin{align*}
	&\braket{\hat{S}^z}_{t_1,\tau,0}=-\frac{N}{2}\cos(\overline{\Omega}t_1)\\
	&\braket{\hat{S}^x}_{t_1,\tau,0}=-\frac{N}{2}\sin(\overline{\Omega}t_1)\sin\{[\delta+(N-1)(\overline{C}-\ell)]\tau\}Z^{N-1}\\
	&\braket{\hat{S}^y}_{t_1,\tau,0}=\textcolor{red}{-}\frac{N}{2}\sin(\overline{\Omega}t_1)\cos\{[\delta+(N-1)(\overline{C}-\ell)]\tau\}Z^{N-1}
\end{align*}
with
\begin{align*}
&Z=\sqrt{\cos^2\overline{X}\tau+\cos^2\overline{\Omega}t_1\sin^2\overline{X}\tau}\\
&\tan\ell\tau=\cos\overline{\Omega}t_1\tan\overline{X}\tau.
\end{align*}
If $\overline{X}\tau\ll1$, then the density shift is 
$$
2\pi\Delta\nu=(N-1)(\ell-\overline{C}) \approx(N-1)(\overline{X}\cos\overline{\Omega}t_1-\overline{C}).
$$

\bibliographystyle{apsrev4-1}
\bibliography{ref}
\end{document}